\documentclass[final,number,sort&compress,5p,times,twocolumn]{elsarticle}
\usepackage{lineno}

\usepackage[]{graphicx}
\usepackage{booktabs}
\usepackage[draft]{todonotes}   % notes shown
\usepackage{amssymb}
\usepackage{eurosym}

\usepackage{multicol}

\usepackage{subfigure}
\journal{Nuclear Instruments and Methods in Physics Research Section A}

\usepackage{etoolbox}
\makeatletter
    \patchcmd{\tnotemark}{\ding{73}}{*}{}{\@latex@error{Failed to path \string\tnotemark\space for \string\ding{73}}}
    \patchcmd{\tnotemark}{\ding{73}\ding{73}}{\dag}{}{\@latex@error{Failed to path \string\tnotemark\space for \string\ding{73}\string\ding{73}}}
    \patchcmd{\tnotetext}{\ding{73}}{*}{}{\@latex@error{Failed to path \string\tnotetext\space for \string\ding{73}}}
    \patchcmd{\tnotetext}{\ding{73}\ding{73}}{\dag}{}{\@latex@error{Failed to path \string\tnotetext\space for \string\ding{73}\string\ding{73}}}
\makeatother

\usepackage{etoolbox}
\makeatletter
\patchcmd{\ps@pprintTitle}% <cmd>
  {Preprint submitted to}% <search>
  {Accepted manuscript by}% <replace>
  {}{}% <succes><failure>
\makeatother
\begin{document}

%\begin{linenumbers}

\begin{frontmatter}

\title{A photomultiplier tube test stand and on-site measurements to characterise the performance of Photonis XP3062 photomultiplier tubes at increased background light conditions and lower gain}

\author[kit]{J.~Zorn\corref{cor}\fnref{fn1}}
\ead{justus.zorn@mpi-hd.mpg.de}
\author[kit]{K.~Daumiller}
\author[kit]{R.~Engel}
\author[kit]{H.-J.~Mathes}
\author[kit]{M.~Riegel}
\author[kit]{R.~\v{S}m\'{\i}da\fnref{fn2}}
\author[kit]{F.~Werner\fnref{fn1}}
%\address[mpik]{now at: Max-Planck-Institut f\"{u}̈r Kernphysik, P.O. Box 103980, 69029 Heidelberg, Germany}
\address[kit]{Karlsruhe Institut f\"{u}̈r Technologie, Kaiserstra\ss{}e 12, 76131 Karlsruhe, Germany}
\cortext[cor]{Corresponding author}

\fntext[fn1]{now at: Max-Planck-Institut f\"{u}̈r Kernphysik, P.O. Box 103980, 69029 Heidelberg, Germany}
\fntext[fn2]{now at: Enrico Fermi Institute and Kavli Institute for Cosmological Physics at the University of Chicago, 5640 South Ellis Avenue,
Chicago, IL 60637, USA }

\begin{abstract}
%% abstract.tex %%
%
%----------------------------------------
%

Photomultiplier tubes (PMTs) are widely used in astroparticle physics experiments to detect light flashes (e.g.~fluorescence or Cherenkov light) from extensive air showers (EASs) initiated by statistically rare very high energy cosmic particles when travelling through the atmosphere. Their high amplification factor (gain) allows the detection of very low photon fluxes down to single photons. At the same time this sensitivity causes the gain and signal-to-noise ratio to decrease with collected charge over the lifetime of the PMT (referred to as ``ageing''). To avoid fast ageing, many experiments limit the PMT operation to reasonably low night sky background (NSB) conditions. However, in order to collect more event statistics at the highest energies, it is desirable to extend the measurement cycle into (part of) nights with higher NSB levels.
%However, to collect more statistics, it is desirable to operate PMTs under high night sky background (NSB) conditionsespecially in astroparticle physics experiments using PMTs to detect light flashes from statistically rare very high energy cosmic particles initiating extensive air showers (EASs) in the atmosphere (like fluorescence or Cherenkov light), operating them also under high night sky background (NSB) conditions is desirable to collect more statistics.
In case the signal-to-noise ratio remains large enough in the subsequent reconstruction of the EAS events, lowering the PMT gain in such conditions can be an option to avoid faster ageing. In this paper, performance studies under high NSB with Photonis XP3062 PMTs, as used in the fluorescence detector of the Pierre Auger Observatory, are presented. The results suggest that lowering the PMT gain by a factor of 10 while increasing the NSB level by a similar factor does not significantly affect the PMT performance and ageing behaviour so that detection and offline reconstruction of EASs are still possible. Adjusting the PMT gain according to a changing NSB level throughout a night has been shown to be possible and it follows a predictable behaviour. This allows to extend the measurement cycles of experiments, based on PMTs of type Photonis XP3062 or comparable and exposed to the NSB, to enhance the sensitivity especially at the highest energies where events are very rare.
\end{abstract}


\end{frontmatter}

\section{Introduction}
\label{introduction}
%% intro.tex %%
%
%----------------------------------------
%

Due to their high sensitivity in detecting low photon fluxes (down to the single photon level), vacuum photomultiplier tubes (PMTs) are widely used in particle and astroparticle physics experiments \citep{Lubsandorzhiev:2006ip}, e.g.~in cameras observing atmospheric fluorescence or Cherenkov light. One of the quantities specifying the operation point of a PMT is its gain $G$ (ratio between cathode and anode current). Since it depends on the high voltage (HV) $U$ applied to the divider network for the individual amplification stages (dynodes), with
\begin{equation}
G\propto U^\alpha
\label{eq:gain-hv}
\end{equation}
and $\alpha$ being a PMT dependent parameter, it can be adjusted according to specific use cases. A decrease in PMT gain and signal-to-noise ratio is observed when they are exposed to a high continuous (background) light while being powered. This effect is referred to as ageing affecting the PMT lifetime which depends on the total amount of accumulated charge on the PMT anode\footnote{While PMT ageing might also be related to other PMT characteristics like total charge removed from the photocathode, in this paper only the dependence on the accumulated anode charge is assessed.}. To avoid fast ageing, in most experiments where PMTs are exposed to night sky background (NSB), the PMT operation is limited to reasonably clear and dark nights which significantly reduces the duty cycle. However, extending the duty cycle is highly desirable for experiments suffering from low event statistics. This could be achieved by observing at higher background light conditions with a reduced PMT gain. For experiments exposed to NSB, this means extending the measurements into nights with a higher moon fraction or into twilight. Such an approach is already investigated and followed by some PMT-based experiments like MAGIC~\citep{Bigongiari:2005sw} and VERITAS~\citep{2002APh....17..221W} where $\gamma$-ray observations are performed also under bright moonlight up to three times the NSB of a dark extragalactic field (as in case of VERITAS~\citep{ARCHAMBAULT201734}) increasing the duty cycle by a factor of $\sim$2 (as in case of MAGIC~\citep{AHNEN201729}). A similar idea has been proposed for the fluorescence detector (FD) of the Pierre Auger Observatory \citep{ThePierreAuger:2015rma} where the duty cycle could be increased by about 50\%, from a current observation period of 15\% to 21\% (reduction due to bad weather, power cuts, and malfunctions already included) \cite{Aab:2016vlz}. However, to go for such an approach, several conditions must be fulfilled: (1) the signal-to-noise ratio does not decrease significantly at lower gain, (2) the PMT performance does not change significantly compared to operation at nominal gain, and (3) the PMT ageing is in an acceptable range in case an exchange of PMTs is not desirable over the lifetime of the experiment. The first condition has already been assessed in Monte-Carlo studies for the Pierre Auger Observatory. They have shown that events at the upper end of the energy spectrum (which are the most desirable due to their extremely low flux) produce enough atmospheric fluorescence photons for a sufficiently high signal-to-noise ratio also at gains being a factor 10 lower than nominal. Conditions (2) and (3) are assessed in this paper.

Measurements to characterise the ageing of the FD PMTs were already executed in the past showing (1) long-term gain drops depending on the total accumulated anode charge with half-lives of 200~C or less and (2) short-term gain shifts depending on the level of illumination and time a PMT was not exposed to a continuous background light \citep{Cla09-publication}. However, all investigations were performed at an initial gain of $G=50\,000$ (referred to as ``nominal FD gain''). The influence of a lower initial gain on the ageing process and on the response of the Photonis XP3062 PMTs (referred to hereinafter as ``FD PMTs'') to light flashes as expected from atmospheric fluorescence or Cherenkov light as well as on their dynamic range and linearity were not explored in those studies.

To investigate those influences, a PMT test stand was designed and built at the Karlsruhe Institute for Technology (KIT) to investigate the ageing and performance of the FD PMTs at lower gain settings to prepare and test the feasibility of a possible extension of the FD duty cycle in the future. The test stand as well as the results from lab and on-site tests with the FD characterising the PMT ageing and performance under increased background light and lower gain are presented in this paper.

\section{Photomultiplier test stand and data analysis}
\label{teststand}
\begin{figure*}[tb]
\centering
\includegraphics[width=1.0 \textwidth]{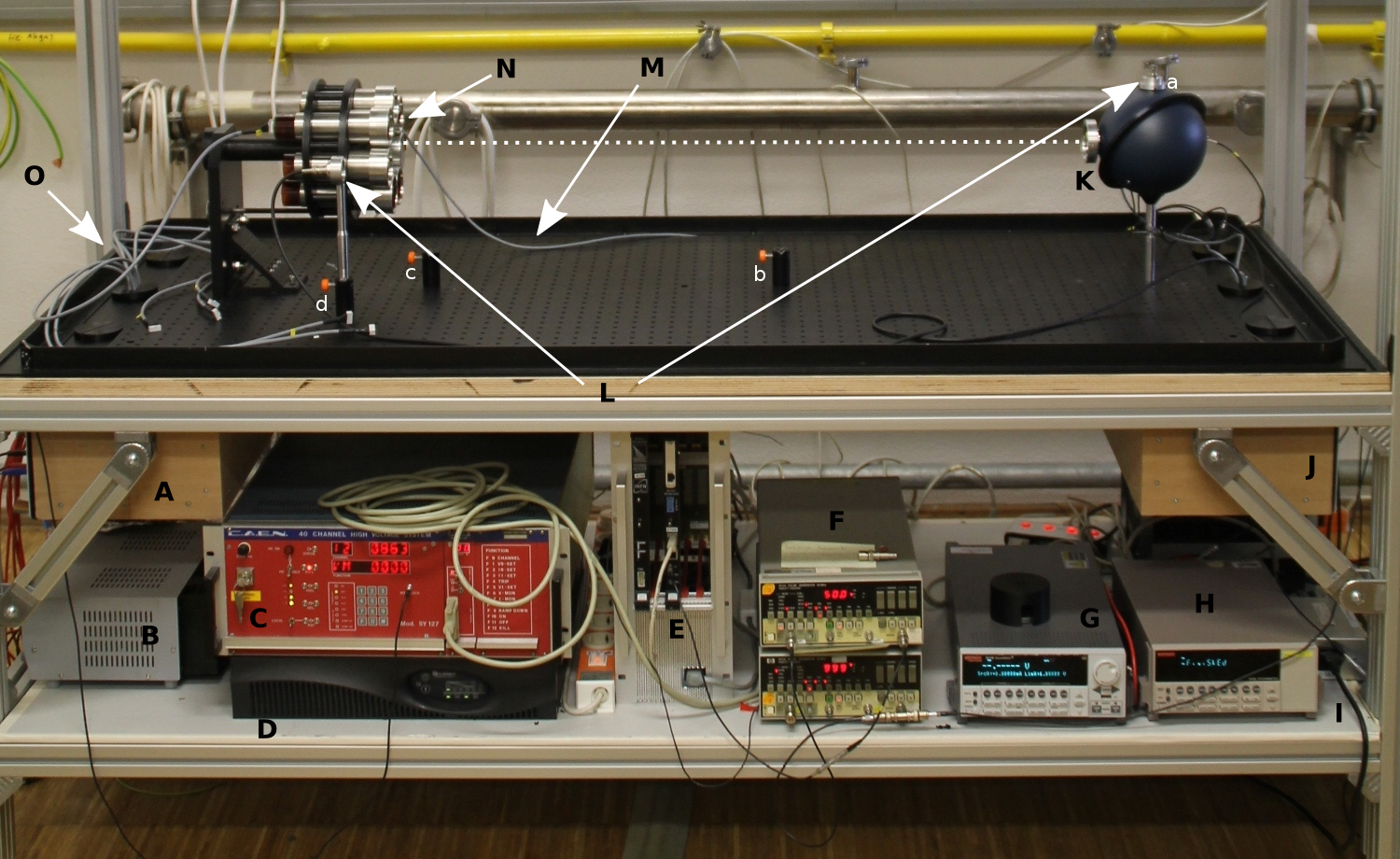}
\caption[]{Detailed view of the (open) PMT test stand. The abbreviations as well as further explanations are presented in the text. \textbf{A}: Light-tight box with cable feed through containing high-voltage distribution board, \textbf{B}: Low-voltage power supply, \textbf{C}: High-voltage crate, \textbf{D}: UPS, \textbf{E}: HEAT data acquisition crate, \textbf{F}: Trigger units, \textbf{G}: SMU, \textbf{H}: Picoammeter, \textbf{I}: Flasher power supply, \textbf{J}: Light-tight box with flasher board and temperature sensor read-out, \textbf{K}: Integrating sphere, \textbf{L}: Monitoring photodiodes (here at position a and d; the other two defined and also marked positions are b and c), \textbf{M}: Temperature sensor, \textbf{N}: PMT wheel with PMTs and Mu-metal tubes, \textbf{O}: Cables for high and low voltage and data transmission. The distance between the exit port of the integrating sphere and the PMTs (Mu-metal entrance) is 97.5~cm (dashed line).}
\label{fig:teststand_detail}
\end{figure*}

%% teststand.tex %%
%
%----------------------------------------
%
The PMT test stand was designed to
\begin{itemize}
\item measure seven PMTs at different HV levels synchronously,
\item illuminate PMTs with continuous and pulsed light simultaneously to emulate real measurement conditions, i.e.~NSB and atmospheric fluorescence or Cherenkov light of EASs, and
\item to allow the extrapolation of the obtained results to on-site observations by using the same readout system as in the experiment.
\end{itemize}
A close-up picture of the test stand is shown in Fig.~\ref{fig:teststand_detail}.
The enclosure is made out of light-tight and lightweight Dibond, a composite panel out of aluminium and polyethylene. It can be closed and opened using a small motor lift. On a $\sim$1.50~m $\times$ 0.70~m plate, the light source, light detectors under test, and auxiliary devices (temperature sensors, photodiodes, etc.) may be mounted in a reproducible fashion.

\begin{figure}[tb]
\centering
\subfigure[]{\includegraphics[width=0.23 \textwidth]{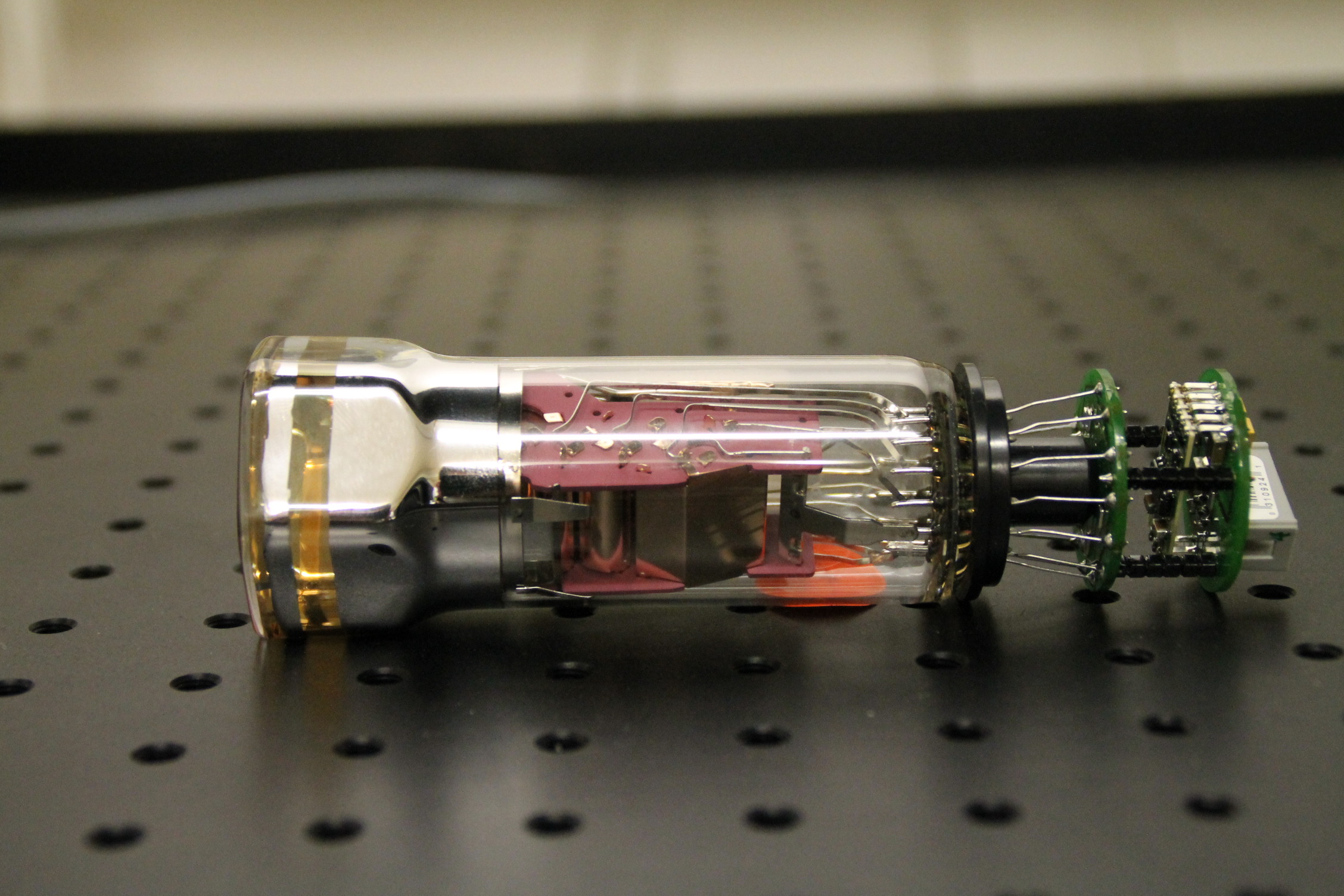}\label{fig:foto_pmt}}
\subfigure[]{\includegraphics[width=0.23 \textwidth]{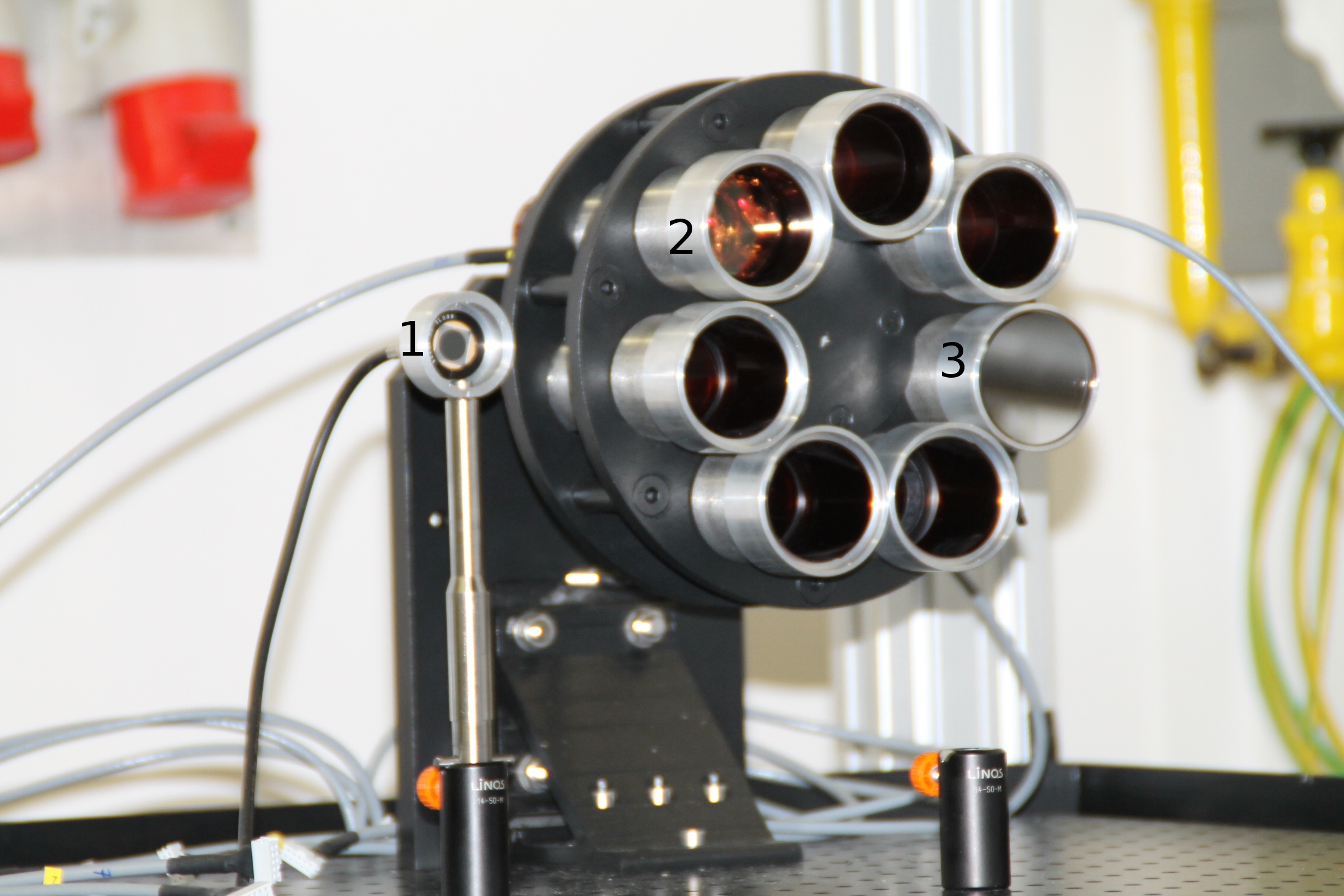}\label{fig:pmt_wheel}} \\
\subfigure[]{\includegraphics[width=0.23 \textwidth]{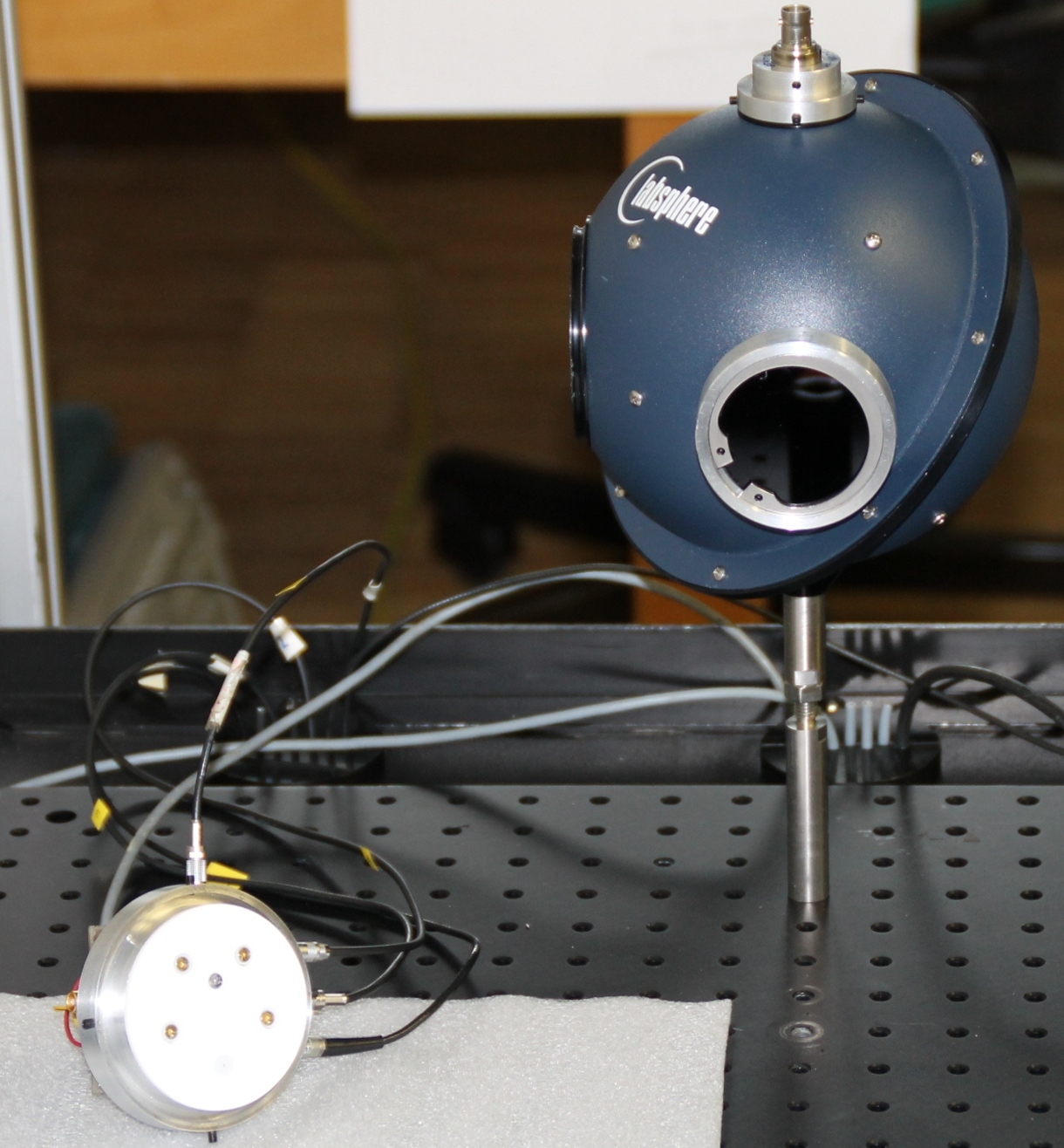}\label{fig:ulbricht}}
\subfigure[]{\includegraphics[width=0.23 \textwidth]{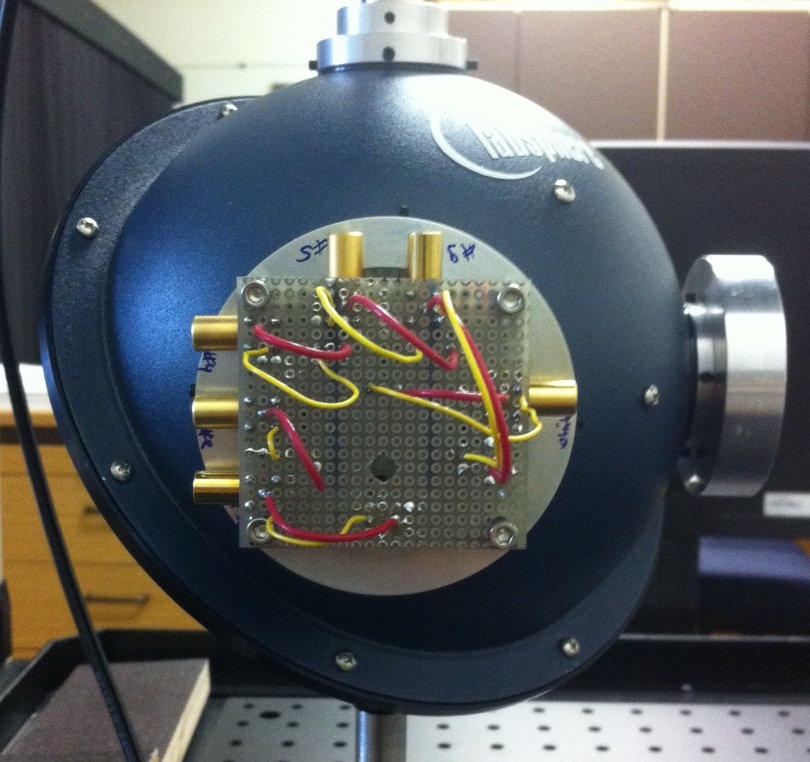}\label{fig:led_plate}}
\caption[]{(a) Photo showing the interior of the Photonis XP3062 PMT with its eight dynodes connected via wires to the head electronics (two green circular planes). (b) Photo showing the PMT wheel with seven Mu-metal tubes serving as positions for seven PMTs. On this photo, two PMTs are mounted (2) \& (3). In front of one PMT (3), an additional neutral density filter is installed. Furthermore, a photodiode (1) is mounted at position d (ref.~Fig.~\ref{fig:teststand_detail}). (c) Photo of Ulbricht sphere with open input port. The interior of the plug is covered by a high-reflective material (Spectralon) like all inner walls of the sphere. The soldered LEDs are flush fitting with the inner surface of the port. A Schott UV bandpass filter is mounted in front of the output port to absorb light with wavelength $>$400~nm. On the upper input port, a photodiode with BNC connector is mounted. (d) Photo of plate with soldered LEDs and Lemo connectors mounted on the input port of the Ulbricht sphere.}
\end{figure}

In the case of the study presented in this paper, the light detectors were nine FD PMTs (see Fig.\ref{fig:foto_pmt}) of type Photonis XP3062 \citep{Pho19,Abraham:2009pm}. They were spare PMTs (stored in dark and temperature-controlled environment) provided by the Pierre Auger Collaboration and had only been used in qualification tests before \citep{2007NIMPA.576..301B}.
They are 1.5 inch hexagonal tubes with a lime glass window and consist of eight linear-focused dynodes. The bi-alkali photocathode is grounded (referred to as positive polarity) to reduce the risk of dust deposition from the ground (when used in the FD) due to electrostatics and to avoid insulating the tube from its surroundings. A capacitor is used to isolate the measuring circuits from the HV, i.e.~the readout is AC-coupled. In such a case, the DC anode current being proportional to the NSB level $f_{\rm{NSB}}$ \footnote{The NSB level is used as a synonym for NSB rate which is defined as the number of NSB photons hitting the PMT window per second.} cannot be measured directly. An indirect method based on Poisson statistics using the signal baseline variance of the measured waveform $\sigma^2$ can be used instead. The relation between $f_{\rm{NSB}}$ and $\sigma^2$ for nominal FD gain is given by (derived from~\citep{Caruso:2005nc,1239269})
\begin{equation}
\frac{f_{\mathrm{NSB}}}{\mathrm{MHz}} \approx 2 \, \frac{A}{\mathrm{m^2}} \, \frac{\Omega}{\mathrm{deg^2}} \, T \, \frac{\sigma^2}{\mathrm{ADC}^2} \approx 7.6 \, \frac{\sigma^2}{\mathrm{ADC}^2},
\label{eq:NSBrate-variance}
\end{equation}
using the FD aperture $A$ of $\sim$3.8~m$^2$, the PMT solid angle $\Omega$ of $\sim$2.0~deg$^2$, and the total optical transmission of the FD system $T$ of $\sim$0.5. At the same time, it holds that
\begin{equation}
f_{\mathrm{NSB}} = \frac{I_{\mathrm{A}}}{\rho \, e \, G},
\end{equation}
where $e$ is the elementary charge, $I_{\rm{A}}$ the anode current, $\rho$ the quantum efficiency, and $G$ the gain of the PMT. Using the nominal FD gain and the quantum efficiency $\rho$ of $\sim$29\% at 375~nm \cite{Kampert:2015lcb}, this results in 
\begin{equation}
\frac{f_{\mathrm{NSB}}}{\mathrm{MHz}} \approx 423.2 \, \frac{I_{\mathrm{A}}}{\mu \mathrm{A}}.
\label{eq:NSBrate-anodecurrent}
\end{equation}

The test stand contains a manually rotatable wheel (see Fig.\ref{fig:pmt_wheel}) to mount seven PMTs together so that all have the same radial distance to the optical axis. Kapton foil and Mu-metal tubes are used for electromagnetic shielding to reduce noise pick-up from other electromagnetic sources in the lab and to reduce position sensitivity in the geomagnetic field. Furthermore, UV-VIS neutral density (ND) filters with different optical densities can be mounted in front of a single PMT to reduce the light flux if necessary.

The light source consists of an Ulbricht sphere (see Fig.~\ref{fig:ulbricht}) from Labsphere, model 3P-GPS-053-SL, and of light-emitting diodes (LEDs) from Roithner LaserTechnik GmbH, type UVLED 370-110E, with peak wavelengths measured to lie around 374~nm at room temperature. The LEDs are soldered to a circuit board mounted to the input port of the Ulbricht sphere (see Fig.~\ref{fig:led_plate}). They can be used either as a continuous DC light source, driven by a source measurement unit (SMU) to emulate the NSB, or as a pulsed light source driven by a custom flasher board, producing pulses with settable width of 2 -- 20~$\mu$s length to emulate fluorescence light from EASs and on-site calibration runs. The flasher board is adapted from the one originally designed for FD calibration measurements with an airborn and remotely controlled light source \citep{Tom16}. A Schott MUG-6 UV bandpass filter, absorbing most parts of the NSB with wavelength $>$400~nm, is mounted in front of the output port of the Ulbricht sphere\footnote{The same type of filter is used at the FD aperture.}.

The distance between the light source and PMT wheel (97.5~cm) was chosen based on an optimisation of light uniformity and intensity of continuous and flashed light at the PMT wheel. As reported by the manufacturer of the Ulbricht sphere, no deviation from uniformity could be measured at the PMT positions, i.e.~at a radial distance of 13~cm from the optical axis (defined by the line between the Ulbricht sphere output port and the PMT wheel centre). The light flux, measured at the PMT positions, can be set in the range of 10 and 2000~photons/50~ns (0.2 -- 40~GHz) for the continuous background light and in the range of 1.5$\times$10$^5$ and 4.5$\times$10$^6$ photons for a 14~$\mu$s light flash.

Silicon photodiodes of Hamamatsu model S2281-01 are used to monitor the continuous background light during measurements and for calibration purposes of the flasher and the LEDs. Four mounting positions (labelled as a, b, c and d in Fig.~\ref{fig:teststand_detail}) were defined. To be able to determine absolute light fluxes, the photodiodes themselves were cross-calibrated with an NIST (US-National Institute of Standards and Technology) calibrated photodiode.

Other devices shown in Fig.~\ref{fig:teststand_detail} are units for measurement control and data read-out:
\begin{itemize}
\item a Voltcraft laboratory power supply PS-2403D for powering the head electronics of the PMTs,
\item a C.A.E.N.~HV system SY127 for HV supply of the PMTs,
\item an uninterrupted power supply (UPS),
\item a HEAT (High-Elevation Auger Telescope \citep{2011ASTRA...7..183M}) data acquisition crate providing full waveform read-out with 20~MHz sampling and 12 bit resolution,
\item two HP 8112 trigger units to trigger the HEAT data acquisition crate and the flasher board,
\item a Keithley SMU 2612B for voltage and current supply of the LED for continuous illumination,
\item a Keithley picoammeter 6485 for accurate measurements of the photodiode currents,
\item several temperature sensors of type DS18B20 for temperature monitoring, and
\item a desktop PC for test stand control and data storage.
\end{itemize}

Before being used in the ageing and performance measurements, the PMTs were characterised (gain-HV dependence, dynamic range, nominal HV determination, etc.) and all other devices were calibrated and tested on their functionality and stability.

In the analysis, the anode charge and gain was determined using the pulse area as charge extraction mechanism. Because of a measured time jitter of the flasher pulse in the order of 1.5--2~$\mu$s (30 to 40 bins in PMT the readout) between consecutive events, the pulse position in the waveform has to be determined first. A pulse in the waveform is being identified in case in $n$ consecutive time bins
\begin{equation}
A_{\rm{ADC}} > \overline{A_{\rm{ADC}}} + 5\cdot \sigma_{\overline{A_{\rm{ADC}}}},
\end{equation}
where $A_{\rm{ADC}}$ is the signal amplitude per time bin in ADC counts, $\overline{A_{\rm{ADC}}}$ the mean signal amplitude of the trace in the time bins before the pulse, $\sigma_{\overline{A_{\rm{ADC}}}}$ the standard deviation of the baseline, and $n$ defined by the programmed flasher pulse length. The samples are then integrated in a given time window determined by the programmed flasher pulse length taking into account PMT afterpulsing and correcting for the baseline skew due to the high-pass filter in the signal chain. Furthermore, the variance of the PMT signal baseline was computed and recorded for every event to monitor its change as a function of background light level and gain.

\section{Ageing of photomultipliers at lower gain}
\label{aging}
\begin{figure}[tb]
\centering
\subfigure[]{\includegraphics[width=0.23 \textwidth]{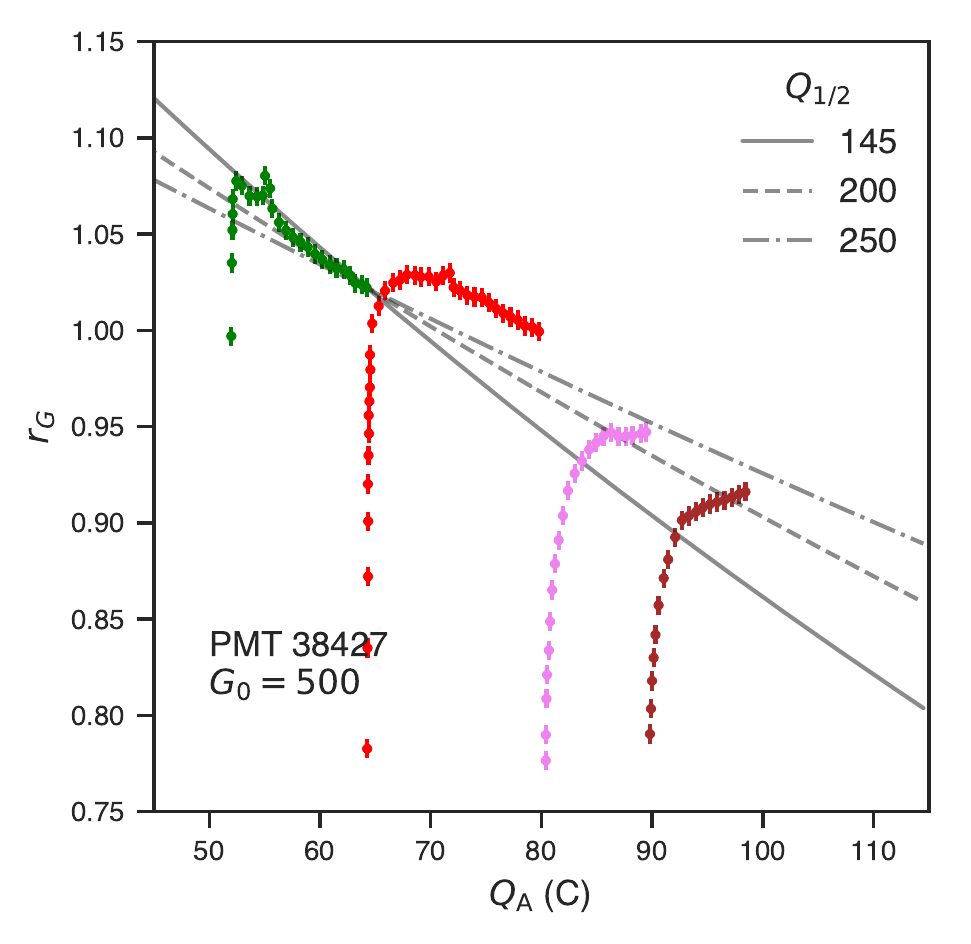}}
\subfigure[]{\includegraphics[width=0.23 \textwidth]{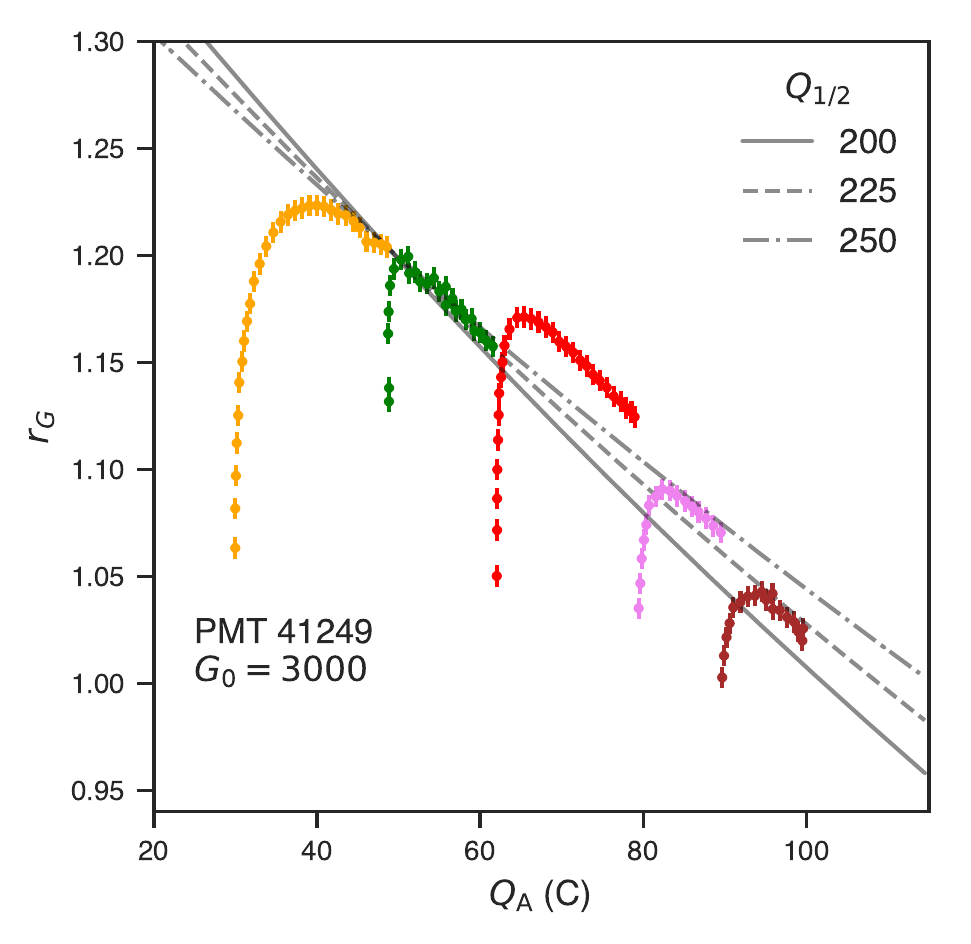}}\\
\subfigure[]{\includegraphics[width=0.23 \textwidth]{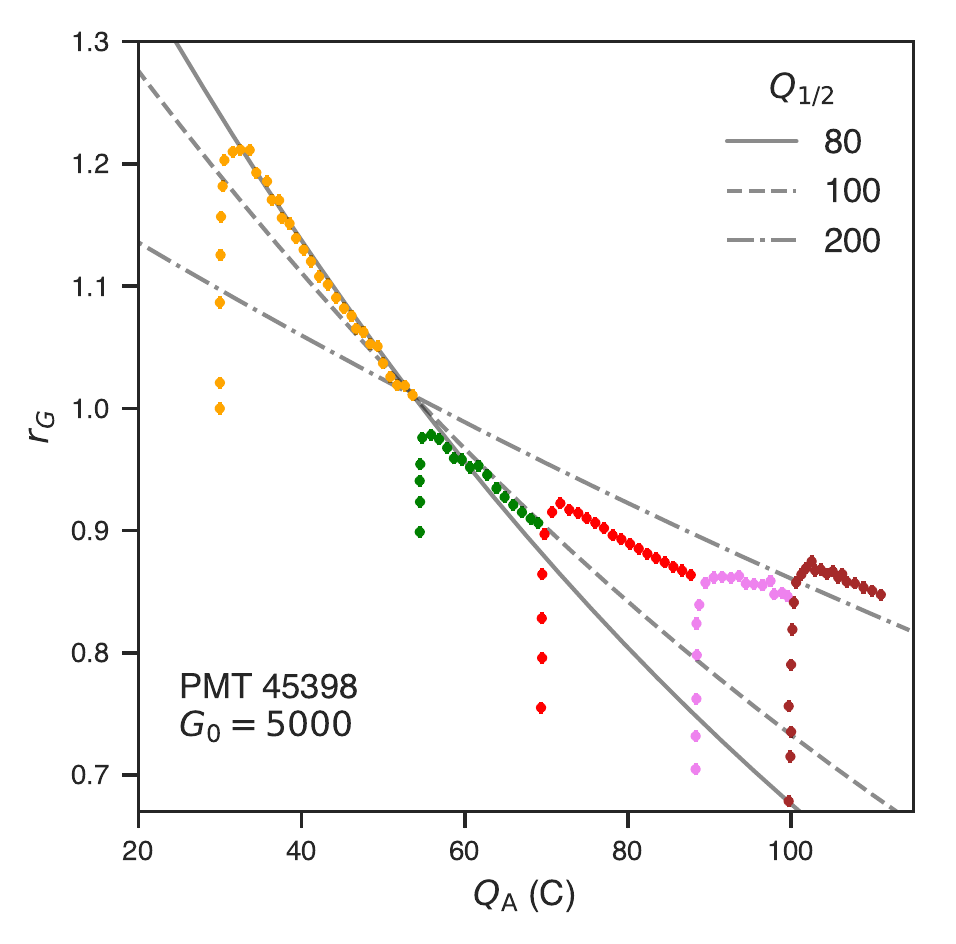}}
\subfigure[]{\includegraphics[width=0.23 \textwidth]{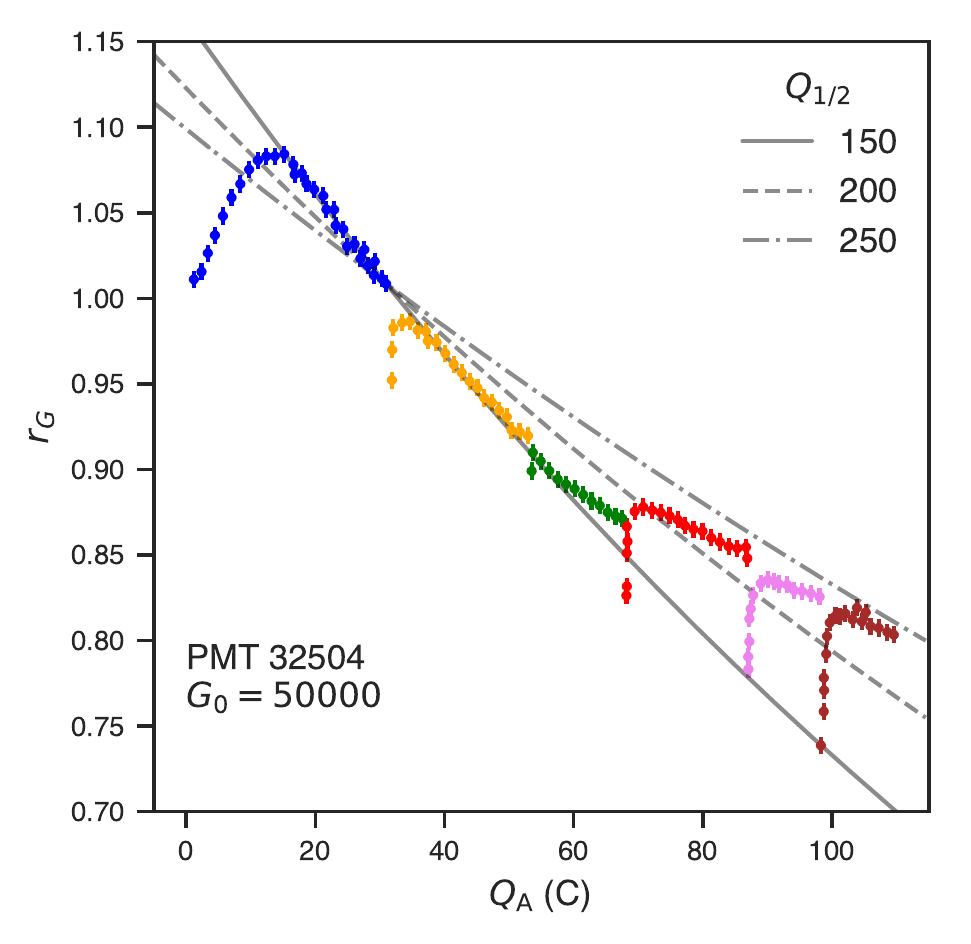}}
\caption[]{Extracted charge ratio $r_G$ (for definition refer to the text) as a function of accumulated anode charge for four different PMTs with different initial gains $G_0$. Each data point represents the mean out of 25 measurements, the errorbar indicates the standard deviation. The different colours correspond to periods of different length of illumination and subsequent break (cf.~Tab.~\ref{tab:periods}). In each case, three exponential functions with different half lives are shown (no fits).}
\label{fig:aging_period3_differentfunctions}
\end{figure}

%% aging.tex %%
%
%----------------------------------------
%
To study the PMT ageing in a reasonable amount of time, a factor $\sim$40 higher background light level than under nominal conditions was used, inducing an anode current of 40~$\mu$A -- the limit up to which the PMT response was measured to be linear. The initial gain of the PMTs used for ageing measurement and mounted in the wheel and the ND filters mounted in front of them were matched to each other. This means that assuming similar ageing behaviour all PMTs were expected to accumulate approximately the same amount of anode charge during the time of the measurement. To investigate also the response of the PMTs to periods with HV and background light being off, the measurements were interrupted by breaks of different lengths. The length of those breaks and of the illumination periods are given in Tab.~\ref{tab:periods}.
\begin{table}[tb]
\caption[]{Duration of the six different periods in the ageing measurements described in the text. The period index, the length of the period the continuous background (DC) light was switched on ($t_{\rm{DC}}$), and the length of the break following a measurement ($t_{\rm{break}}$), both in units of days (d), are recorded in the table.}
\vspace{0.5cm}
\centering
\begin{tabular}{cccc}
\toprule 
Period & Data point colour in Fig.\ref{fig:aging_period3_differentfunctions} & $t_{\rm{DC}}$ (d)& $t_{\rm{break}}$ (d)\\ 
\midrule
1 & \color{blue}{blue} & 6.60 & 1.08\\
2 & \color{orange}{orange} & 5.50 & 0.16\\
3 & \color{green}{green} & 3.50 & 2.25\\
4 & \color{red}{red} & 5.00 & 3.80\\
5 & \color{pink}{pink} & 3.25 & 3.80\\
6 & \color{brown}{brown} & 3.25 & \\
\bottomrule
\end{tabular}
\label{tab:periods}
\end{table}
The response of the PMTs was monitored using a set of 50 pulsed light flashes with a frequency of 1~Hz emitted every 10 minutes. The ratio $r_G$ of the extracted charge at measurement $n$ and the first measurement is calculated to track the gain decrease with accumulated anode charge $Q_{\rm{A}}$. $Q_{\rm{A}}$ is calculated for each measurement $n$ and PMT using
\begin{itemize}
\item the LED photon flux $\Phi_\gamma$ at the PMT window, measured by the photodiode at position d (cf.~Fig.~\ref{fig:teststand_detail}) and taking into account the ND filter attenuation,
\item the PMT quantum efficiency $\rho$, assumed to be the same for each PMT and constant over time,
\item the initial gain $G_0$,
\item the gain decrease determined by $r_G$,
\item the integration time $\Delta t = t_n - t_0$ of the photodiode measurement, and
\item the elementary charge $e$.
\end{itemize}
This leads to
\begin{equation}
Q_{\rm{A}} = \int_{t_0}^{t_n} r_G(t)\,G_0\,\rho\,e\,\Phi_\gamma\,dt.
\end{equation}
Previous PMT ageing investigations with a nominal initial gain of $G_0=50\,000$ found that the gain decrease follows an exponential function
\begin{equation}
f(Q_{\rm{A}}) = A\,\exp\left(-\ln 2 \frac{Q_{\rm{A}}}{Q_{1/2}}\right)
\end{equation}
characterised by the half-life $Q_{1/2}$ (cf.~\citep{Cla09-publication}). Fig.~\ref{fig:aging_period3_differentfunctions} shows the gain decrease over the six periods for four different PMTs with different initial gains as function of the accumulated anode charge as well as three exponentials with different half-lives for each PMT.
Although all four PMTs used in the measurements were spare PMTs and had only been used in qualification with a negligible contribution to the accumulated anode charge, the starting point of three of them is not at 0~C (see Tab.~\ref{tab:fit_parameters}): PMTs 41249 and 45398 ran through unstable conditions in the first period and PMT 38427 additionally also in the second period.
\begin{table}[tb]
\caption[]{Half-lives $Q_{1/2}$ for different PMTs, initial gains $G_0$ (adjusted using the known gain-voltage-dependence of each PMT), and ranges of accumulated anode charge $Q_{\rm{A}}$, obtained either by an exponential fit or lower limit estimation (in case the fit underestimates the half-life).}
\vspace{0.5cm}
\centering
\begin{tabular}{ccccc}
\toprule 
PMT & $G_0$ & $Q_{\rm{A,start}}$ (C) & $Q_{\rm{A,end}}$ (C) & $Q_{1/2}$ (C) \\
\midrule
37628 & 50\,000 & 0 & 100 & $>65$\\
37384 & 50\,000 & 0 & 45 & 100\\
 & 5000 & 45 & 80 & 170\\
 & 50\,000 & 80 & 160 & $>170$\\
45248 & 50\,000 & 0 & 45 & 167\\
 & 10\,000 & 45 & 180 & $>167$\\
38093 & 50\,000 & 0 & 30 & 64\\
38427 & 500 & 50 & 100 & 200\\
41249 & 3000 & 30 & 100 & 226\\
45398 & 5000 & 30 & 115 & $>80$\\ 
32504 & 50\,000 & 0 & 110 & $>150$\\
\bottomrule
\end{tabular}
\label{tab:fit_parameters}
\end{table}
Those measurements were excluded from the analysis and the starting point for those PMTs was assumed to be of the order of the accumulated anode charge of PMT 32504 at the beginning of the second and third period, respectively.

It can be observed that
\begin{itemize}
\item PMT 32504 shows a similar significant gain increase as observed in previous measurements at nominal initial gain (cf.~\citep{Cla09-publication}), referred to as initial ageing. The increase is about 10\% up to an accumulated anode charge of $Q_{\rm{A}}\sim$15~C,
\item similar to previous tests (cf.~\citep{Cla09-publication}), a break causes the PMT sensitivity first to drop before recovering back to a similar level as before the break for initial gains $G_0$ of 3000, 5000, and 50000 -- this behaviour is different for PMT 38427 with an initial gain of $G_0 = 500$ where no full recovery is observed,
\item for all PMTs the slope of the exponential function describing the ageing process is different between two periods, the half lives being increasing after each break as can be seen by the different exponentials plotted with the data. A single exponential with only one characteristic half-life seems not to be appropriate to fit the data in those cases. The lowest half-life of all exponentials plotted to the data provides a ``lower limit estimation''.
\end{itemize}

Certain differences between different PMTs (exact influence of breaks, half-lives, etc.) and initial gains can be observed. Especially the influences of breaks on the PMT with a gain of $G=500$, observed to not fully recover back to the ``pre-break-level'' after a break, is different compared to PMTs with a higher initial gain. % Since the operating HV of this PMT to achieve a gain of $G=500$ is 373~V which is way below the operation limits for PMTs of type Photonis XP3062 being 880--1300~V, a different behaviour is expected. 
For all tested PMTs, a sensitivity loss described by the decrease of the extracted charge ratio is observed as function of the accumulated anode charge with half-lives in the order of $\sim$60 to $\sim$170~C (see Tab.~\ref{tab:fit_parameters}). No systematic gain dependence on the value of the half-life has been found.

As mentioned previously, the ageing study presented in this paper investigated the dependence of the gain change on the accumulated anode charge and not on other PMT characteristics like total charge removed from the photocathode. The most likely explanation for ageing is physical damage of the anode and the last dynode caused by the permanent electron bombardment and resulting in changes in their material property and surface state. To characterise the exact (micro-)physical mechanism a deeper investigation would be necessary such as a microscopic investigation of the anode and dynodes of an aged PMT.

\section{Performance of photomultipliers at increased background light and lower gain}
\label{results}
\begin{figure}[tb]
\centering
\includegraphics[width=0.5 \textwidth]{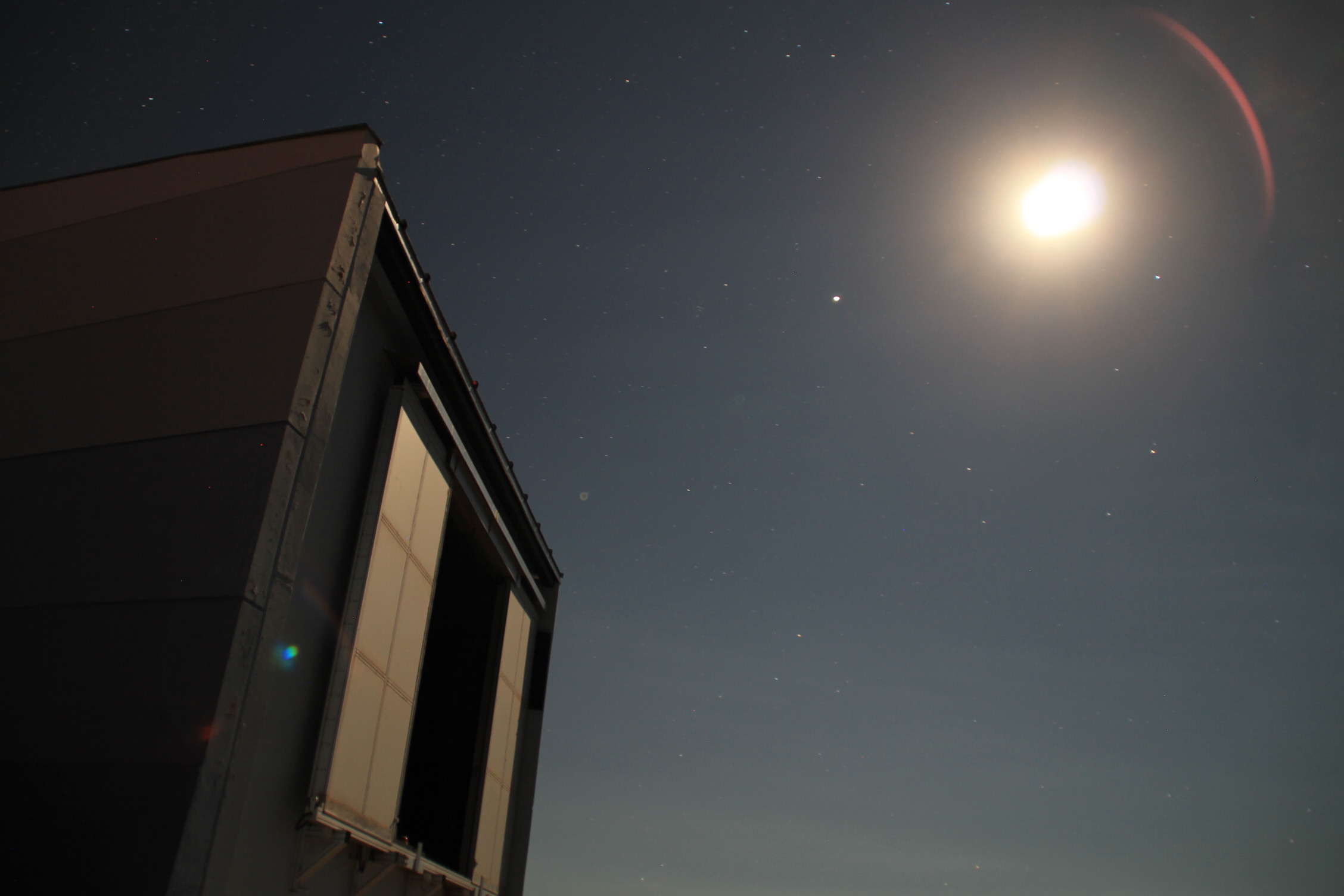}
\caption[]{Photo taken during the sixth test night when the illuminated fraction of the moon was 98\% at 1:00 ART local time. It shows the Moon, Jupiter and the opened shutter of one of the Auger fluorescence telescope Los Leones 1.}
\label{fig:photo_openedshutter}
\end{figure}

%% results.tex %%
%
%----------------------------------------
%
To be able to operate PMT based experiments also at high background light conditions, i.e.~to increase the duty cycle, it is desirable to adjust the PMT gain depending on the background light level such that the performance does not suffer significantly over the foreseen lifetime of the experiment. Fig.~\ref{fig:photo_openedshutter}, showing a picture taken when one of the Auger FD cameras (in bay 1 at the Los Leones site, referred to hereinafter as Los Leones 1) was operated at a gain of $G=3000$, i.e.~at a gain of 16.6 times lower than nominal ($G_{\rm{nom}}=50\,000$), and a moon fraction of 98\%, should give an impression which kind of nights could be added to the nominal schedule if data can be taken at reduced gain.

To investigate the performance at different gain settings in more detail, FD PMTs were operated in the test stand at a varying background light levels emulating the varying NSB and flashed with pulses to monitor the PMT response. Furthermore, to test the performance also under real conditions, Los Leones 1 was operated at different gain settings in five consecutive nights which were not part of the nominal shift schedule since the NSB level exceeded the nominal limits\footnote{To avoid fast ageing of the PMTs, under nominal operation the FD shutters are closed automatically in case (1) $\sigma^2 > 1000$~ADC$^2$ (corresponding to $f_{\rm{NSB}} > 7.6$~GHz) in one single PMT for more than 1.5 minutes or  (2) $\sigma^2 > 100$~ADC$^2$ (corresponding to $f_{\rm{NSB}} > 760$~MHz) in at least 25 PMTs
of the same camera for more than 1.5 minutes.}. The gain settings were chosen from the expected NSB level for the given nights and from results of lab measurements (see below). The exact HV values for a given gain were calculated for each PMT individually using the gain-HV-dependence (cf.~Eq.~\ref{eq:gain-hv}), its nominal HV, and $\alpha$, both stored in a database \citep{2007NIMPA.576..301B}.

Since the PMT signal baseline variance is proportional to the anode current and the NSB level (cf.~Eq.~\ref{eq:NSBrate-variance} \& \ref{eq:NSBrate-anodecurrent}), continuous monitoring of its value delivers detailed information about the current NSB level and the PMT anode current. This feature should remain usable also at lower gain settings so that (1) no additional device is necessary for NSB monitoring and (2) the PMT operational limits (when to close the shutters and when to switch between two gain settings) could still be defined based on the measurement of the signal baseline variance.

In order to verify this and to investigate how a procedure on site could be defined, i.e.~when the PMT gain/HV should be adjusted according to changing NSB conditions (e.g.~twighlight at beginning and end of a night), measurements with a varying background light emulating varying NSB were performed with the test stand.
\begin{figure}[tb]
\centering
\includegraphics[width=0.5 \textwidth]{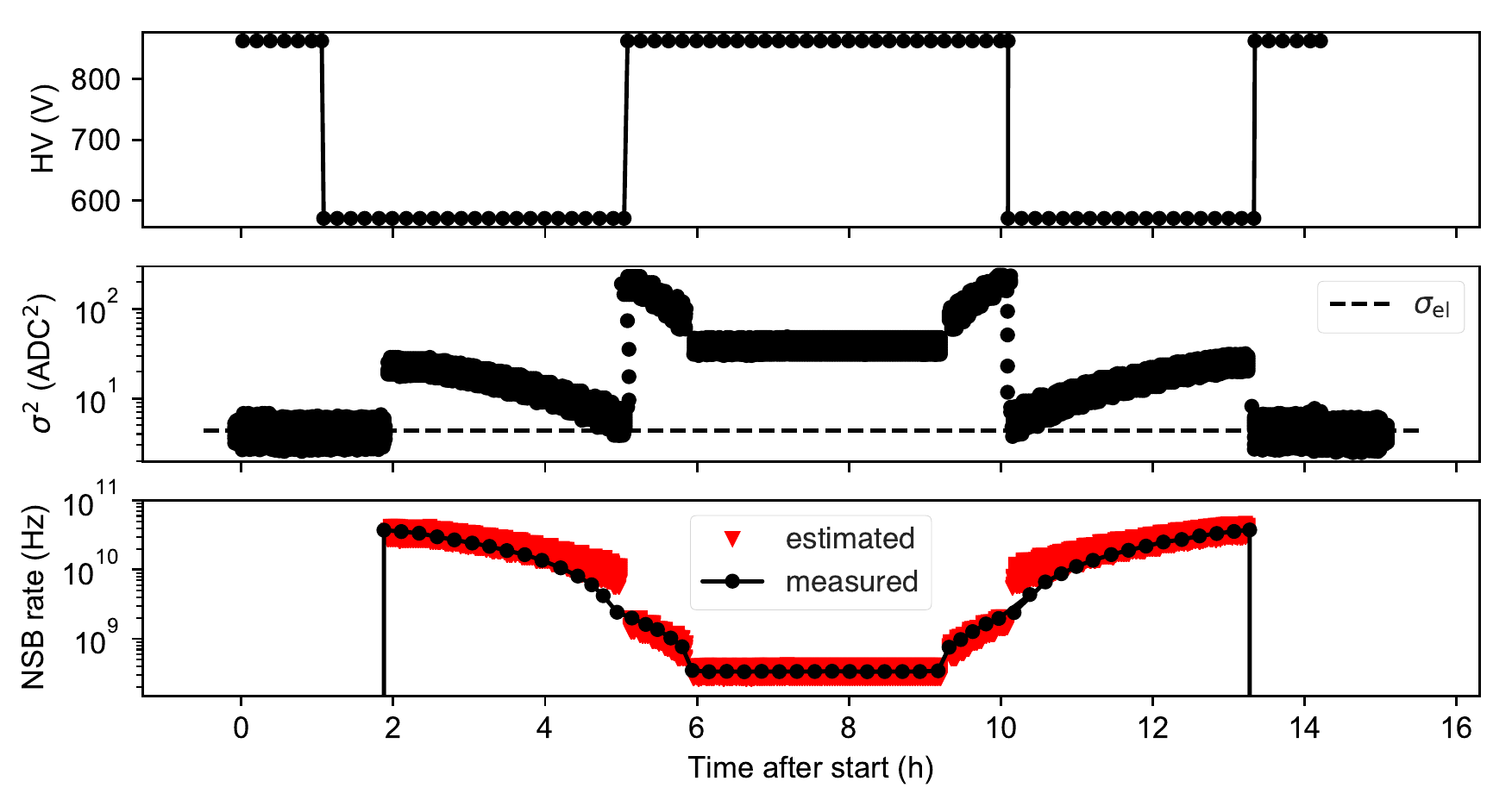}
\caption[]{Emulation of a nightly measurement for PMT 41249 with two gain settings and altering background light conditions. The top panel shows the HV supplied to the PMT. For the higher (lower) gain period, a gain of 50\,000 (4000), corresponding to 863~V (571~V), was used. The middle panel shows the signal baseline variance measured by the PMT while the continuous LED photon flux changed. The bottom panel shows the LED photon rate hitting the PMT window (NSB rate) (a) measured by the photodiode close to the PMT wheel (black data points) and (b) estimated from the PMT signal baseline variance measurements taking into account the PMT gain (red data points). }
\label{fig:emulation_g4000}
\end{figure}
In the measurement shown in Fig.~\ref{fig:emulation_g4000}, the PMT gain was increased/decreased by a factor of 12.5 as soon as the NSB rate fell below/increased above the limit of 2~GHz corresponding to $\sigma^2\sim$ 263~ADC$^2$ at nominal gain (to approximate the on-site limits). The emulated NSB level was measured with one of the photodiodes. However as shown by Fig.~\ref{fig:emulation_g4000} (middle and lower panel), the NSB rate could also be estimated from signal baseline variance measurements at both, nominal and factor 12.5 lower gain. Only when the measured variances due to NSB were in the order of the electronic noise fluctuations ($\sim$4~ADC$^2$), the NSB level estimated from variance measurements deviates from the true NSB level. However, since the measured electronic noise on site is somewhat lower than the one measured in the lab ($\sim$3~ADC$^2$), a gain of $G=4000$ can be seen as a lower limit for the lower gain setting if the condition ``the NSB level should always be able to be estimated by signal baseline variance measurements'' should remain.

\begin{figure}[tb]
\centering
\includegraphics[width=0.5 \textwidth]{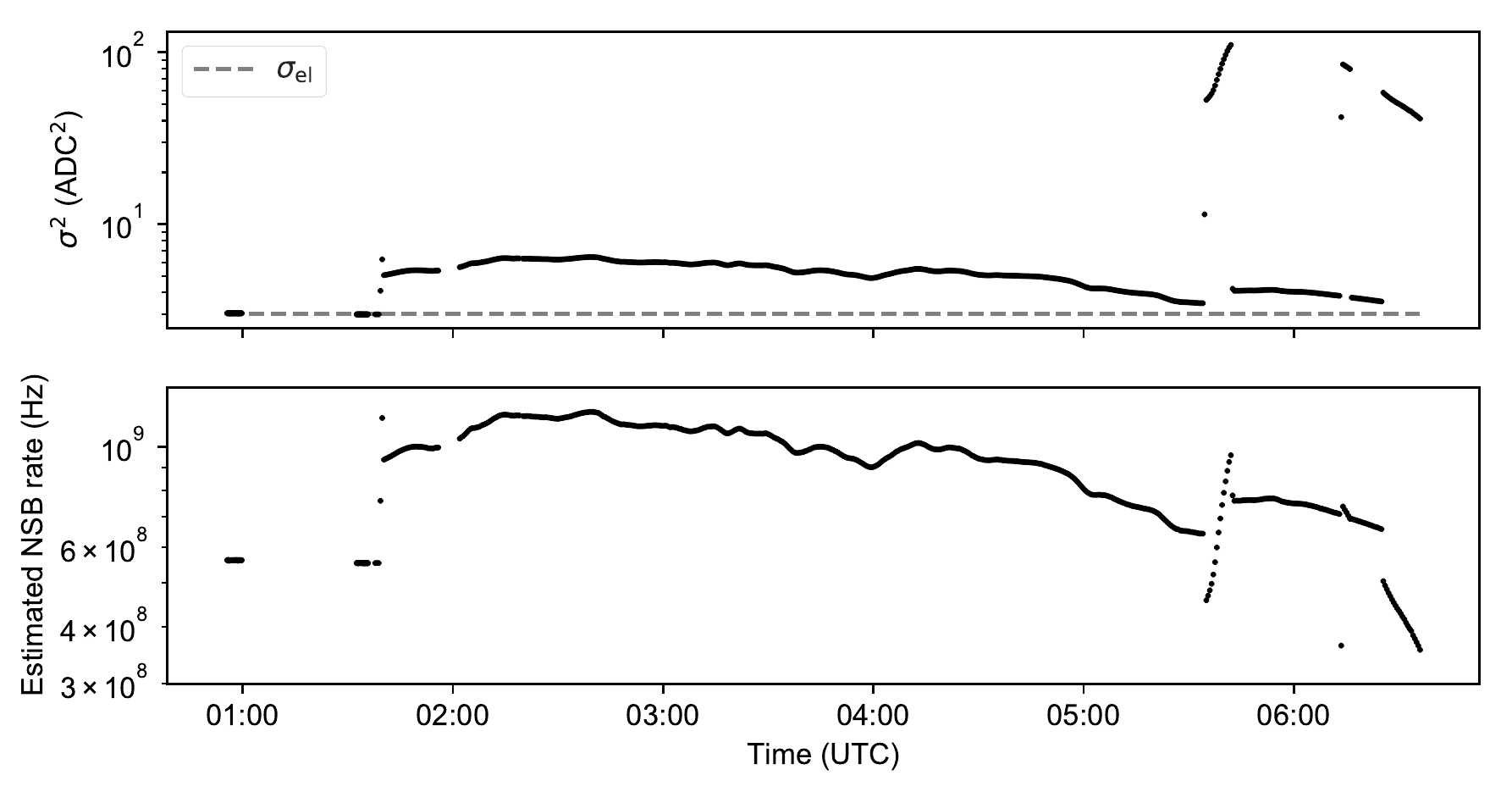}
\caption[]{Average signal baseline variance of all PMTs of Los Leones 1 (top) and the NSB level estimated from signal baseline variance measurements (bottom) as function of time in the night Feb 28th on Mar 1st, 2015. During the first 4 hours of observation, the camera was operated at a gain of $G=5000$. Three times (at 5:33 UTC, 6:13 UTC, and 6:26 UTC) the gain was switched from lower to nominal. Twice (at 5:42 UTC and 6:15 UTC), it was set back to the lower gain after some minutes.}
\label{fig:variance_switching_3rdnight}
\end{figure}

Measurements on site with Los Leones 1 show similar results. Fig.~\ref{fig:variance_switching_3rdnight} shows the signal baseline variance (average out of 352 PMTs\footnote{Only 352 PMTs are present in the camera of Los Leones 1 due to reuse of one fifth of the PMTs for other purposes within the observatory.}) and the calculated NSB rate using the variance measurement throughout a night in which the camera was first operated at an NSB level being 5--10 times higher and a 10 times lower gain than nominal.
Towards the end of the night, we tried to increase the camera gain back to nominal several times since the NSB rate decreased. However, we decided twice to go back to the reduced gain because the variances indicated that the sky was still too bright for nominal operation. Finally, at 6:26 UTC, the sky was dark enough (due to moonset) to operate the camera at nominal gain for the rest of the night. The discontinuities in the estimated NSB rates observed at PMT gain switches are expected to be due to (1) not-instant HV change, i.e.~some PMTs still in transition from lower to higher HV value, and (2) gain spread between pixels in the camera.

\begin{figure}[tb]
\centering
\includegraphics[width=0.5 \textwidth]{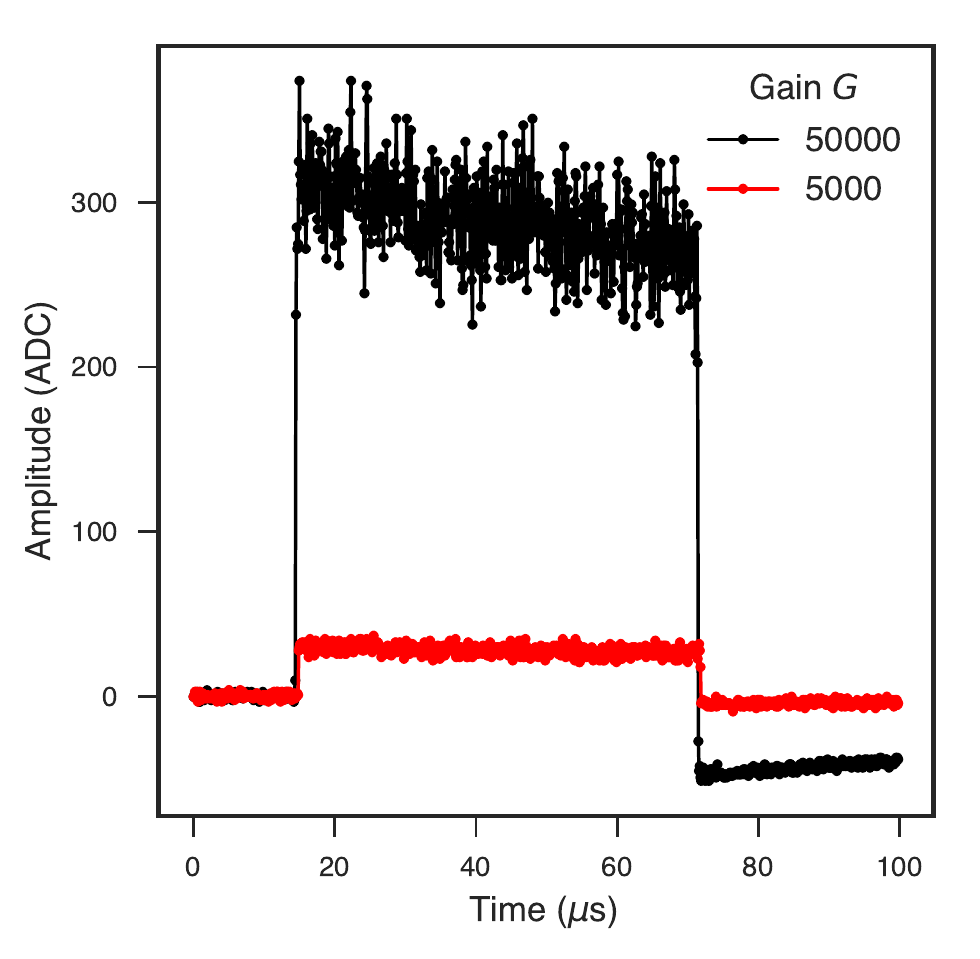}
\caption[]{Calibration flashes recorded by a camera pixel at nominal and lower gain (10 times lower than nominal) with closed shutter. Due to a high-pass filter in the read-out electronics, the baseline drifts in case of a pulse detection causing the signal amplitude to decrease with time and the amplitude being below 0 after the pulse (cf.~Sec.~\ref{teststand}).}
\label{fig:CalAflashes}
\end{figure}

\begin{figure}[tb]
\centering
\includegraphics[width=0.5 \textwidth]{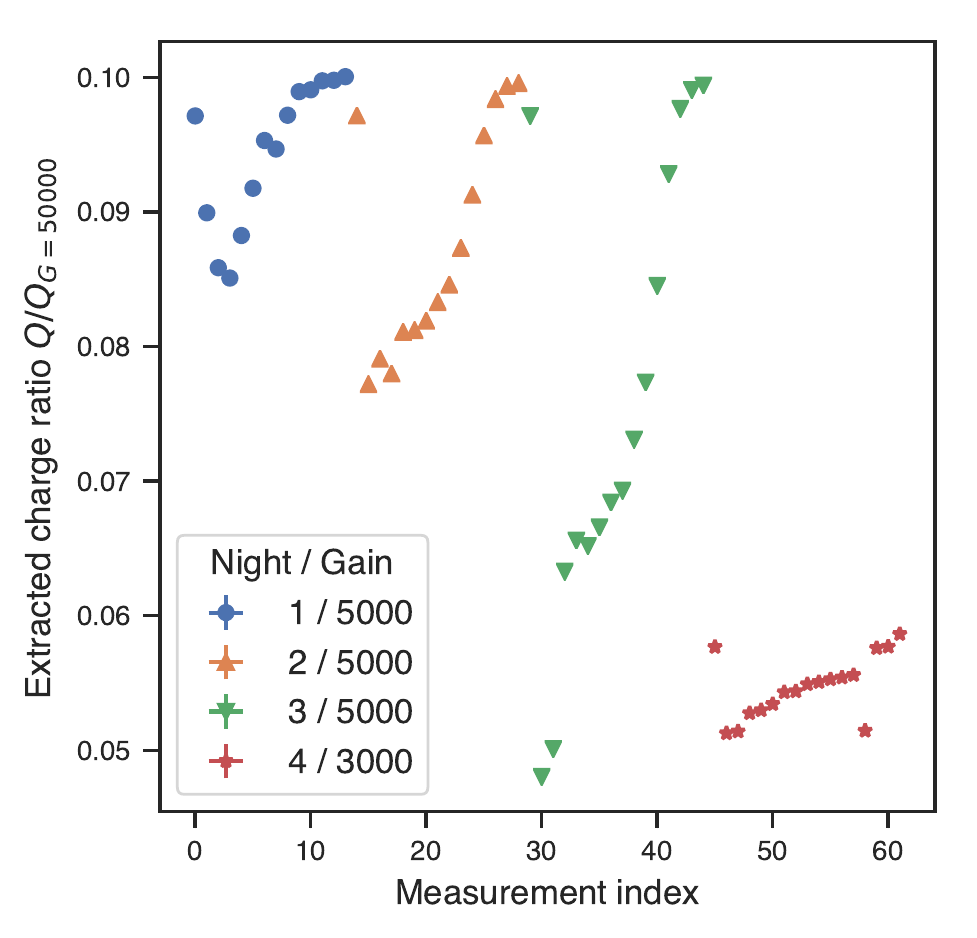}
\caption[]{Relative PMT response (extracted charge ratio, i.e.~charge normalised to the PMT response at nominal gain and closed shutter on the first night) to calibration flashes at lower gains for the mean of the whole camera of the Auger FD telescope Los Leones 1 in five consecutive nights.}
\label{fig:calA}
\end{figure}

In addition to the signal baseline variance, the response of the PMTs to pulses was investigated to evaluate their performance in nights with different gain settings. For this purpose we used the so-called ``Continuous Calibration A'' (cf.~\citep{Tom16}): interleaved measurements illuminating the camera of Los Leones 1 every 30 minutes with 10 flashes of rectangular shape and 60~$\mu$s length with a 470~nm LED while the camera being in operation, i.e.~HV on, shutter open, and camera exposed to NSB. Typical traces of such calibration measurements recorded by one of the camera pixels for two different gain settings ($G_{\rm{nom}}$ and $G=4000$) are shown in Fig.~\ref{fig:CalAflashes}.
As expected, the pulse amplitude ratio between the two gain settings is $\sim$10. Fig.~\ref{fig:calA} shows the relative response of the FD camera when operated at a gain lower than nominal to such calibration measurements in five consecutive nights (same nights as mentioned above).
In each night, the first and last measurement were performed with closed shutter, the others were ``Continuous Calibration A'' measurements.

It can be concluded that
\begin{enumerate}
\item the system is sensitive to calibration pulses also at reduced gain such that the change of the system response throughout a night can be monitored and evaluated also at reduced gain, and
\item the variation of the response throughout a night is similar to what is observed at nominal gain (cf.~\citep{Tom16}).
\end{enumerate}
It is likely that the observed variations result from gain drifts due to exposure of the PMTs to the NSB \citep{Tom16}, more in detail caused by electrical and thermal effects in the PMT and its circuitry \citep{1972ApOpt..11.2576S}. Those effects could be transverse current on dynodes and current flow through the PMT base as well as heating of the anode causing short-term variations of its physical surface properties. To characterise the exact mechanism causing the observed variation of the PMT response, further investigation on a micro-physical level would be necessary which were not within the scope of this work. In any case, the measurement results shown in Fig.~\ref{fig:calA} stress the importance of continuous gain monitoring using interleaved calibration measurements. The study in this paper showed that those measurements are possible at both, nominal and lower gain.  

\begin{figure*}[tb]
\centering
\subfigure[]{\includegraphics[width=0.49 \textwidth]{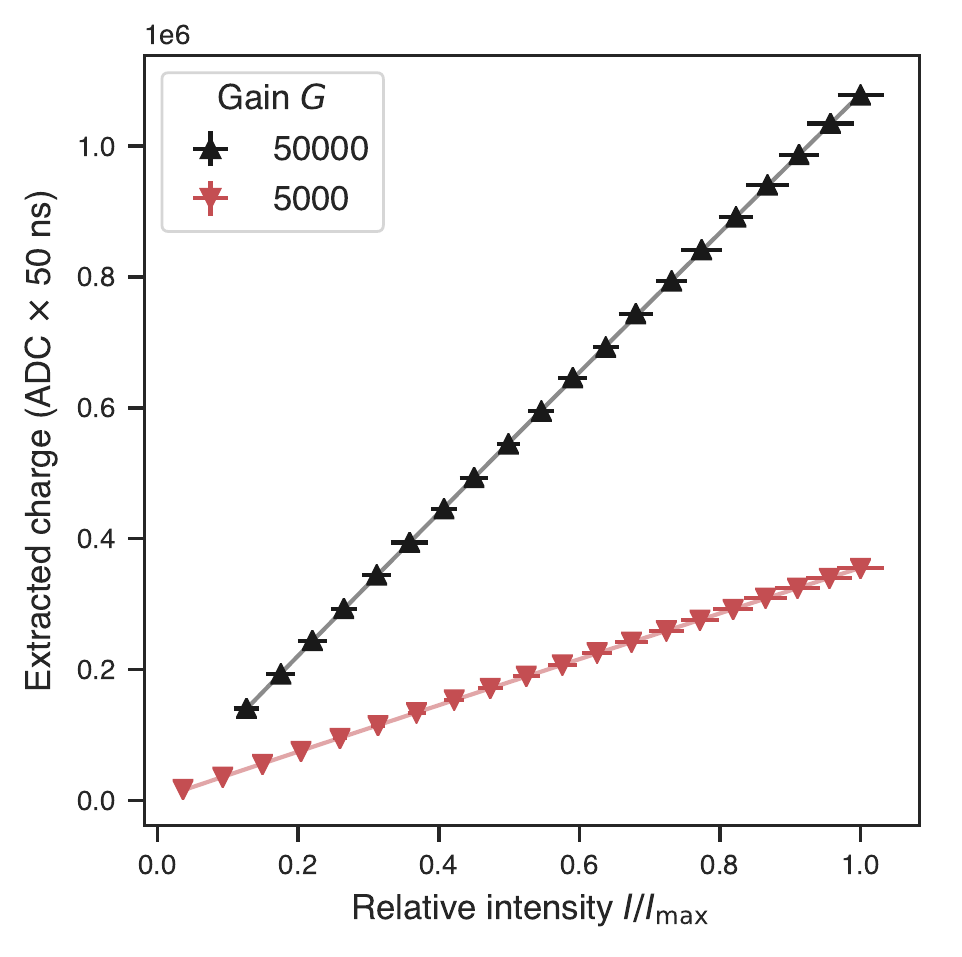}}\label{fig:linearity}
\subfigure[]{\includegraphics[width=0.49 \textwidth]{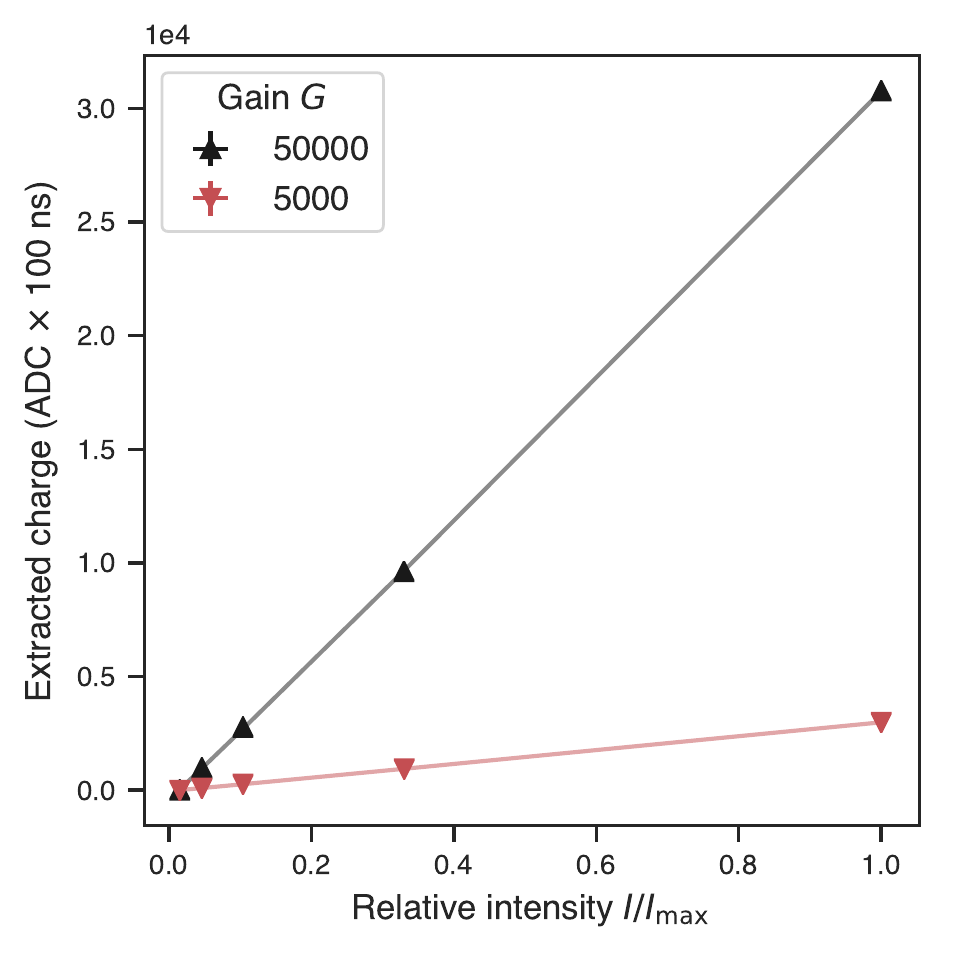}}\label{fig:CalBanalyses}
\caption[]{Response of (a) one Auger FD PMT in the lab and (b) the whole Auger FD camera of the telescope Los Leones 1 (adapted from \citep{Tom16}) to light flashes of different amplitudes as function of the relative light intensity for two different gain settings ($G_{\rm{nom}}=50000$ and $G=5000$). Each data point in (a) represents the mean out of $\sim$20 measurements using the flasher unit at different amplitudes for each gain during three consecutive emulated night shifts (such as shown in Fig.~\ref{fig:emulation_g4000}). The data points in (b) represent the camera mean response (352 pixels) for one measurement using a Xenon flash lamp installed in the telescope (with different pulse shape than the flasher unit pulses) in combination with filters of different transmission coefficients. While in (a) the PMT was exposed to a continuous background light during the dynamic range measurement, the measurement in (b) was executed with closed shutter, thus without background light.}
\label{fig:dynamic_range}
\end{figure*}

Furthermore, the PMT linearity was investigated in the lab and on site. In the lab, the flasher board was used to loop through different amplitudes under the presence of continuous background light. On site, the measurement was performed with closed shutter, i.e.~no background light using a Xenon flash lamp and a filter wheel with five filters with $>$2 orders of magnitude difference in their transmission coefficients (referred to as ``Calibration B'', cf.~\citep{Tom16}). In both cases, the linearity was investigated at two different gain settings: at nominal and a factor 10 reduced gain. Fig.~\ref{fig:dynamic_range} shows the results. The linearity was found to be independent of the gain with a power law coefficient compatible with 1 even when the measurement was done with the PMT being exposed to continuous background light (as in case of the lab measurement).

\begin{figure*}[h!]
\centering
\subfigure[]{\includegraphics[width=0.49 \textwidth]{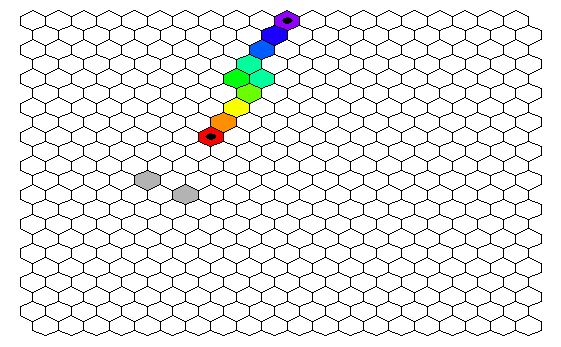}}
\subfigure[]{\includegraphics[width=0.49 \textwidth]{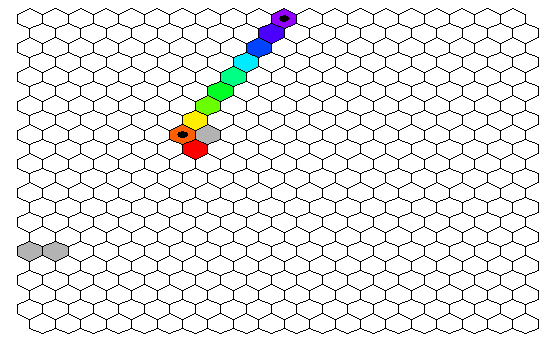}}\\
\subfigure[]{\includegraphics[width=0.49 \textwidth]{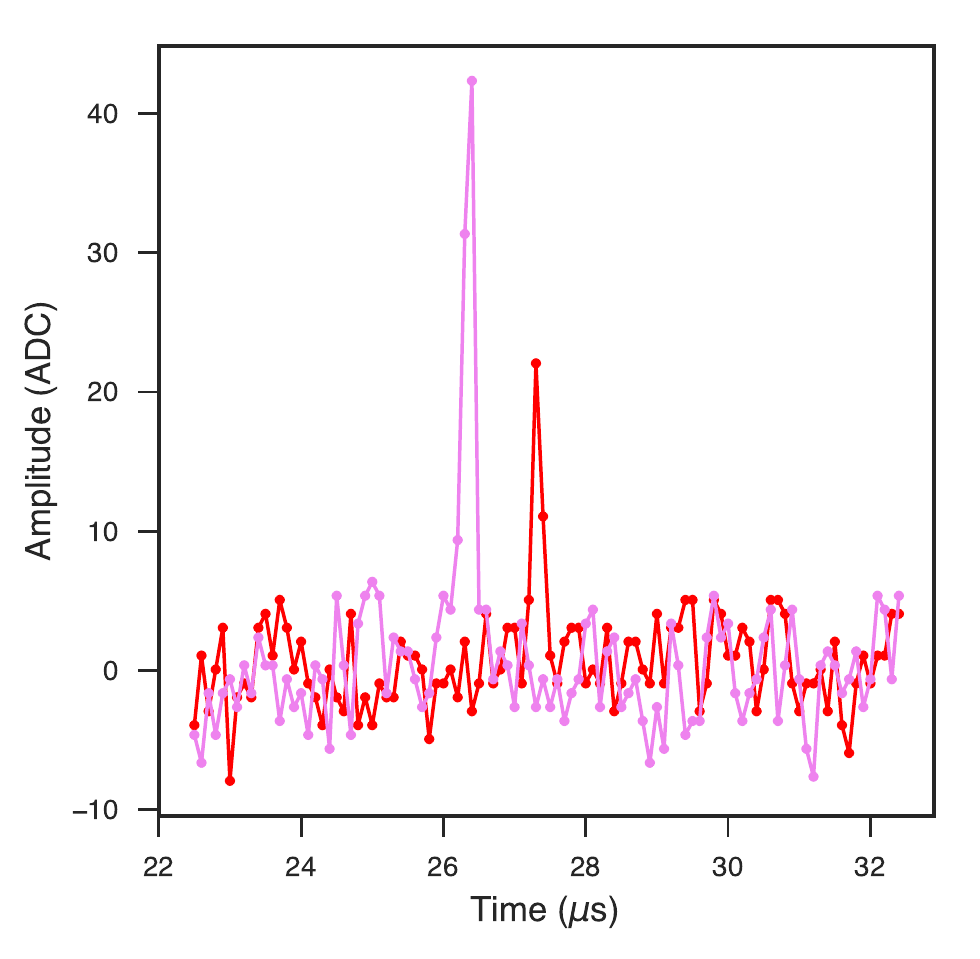}}
\subfigure[]{\includegraphics[width=0.49 \textwidth]{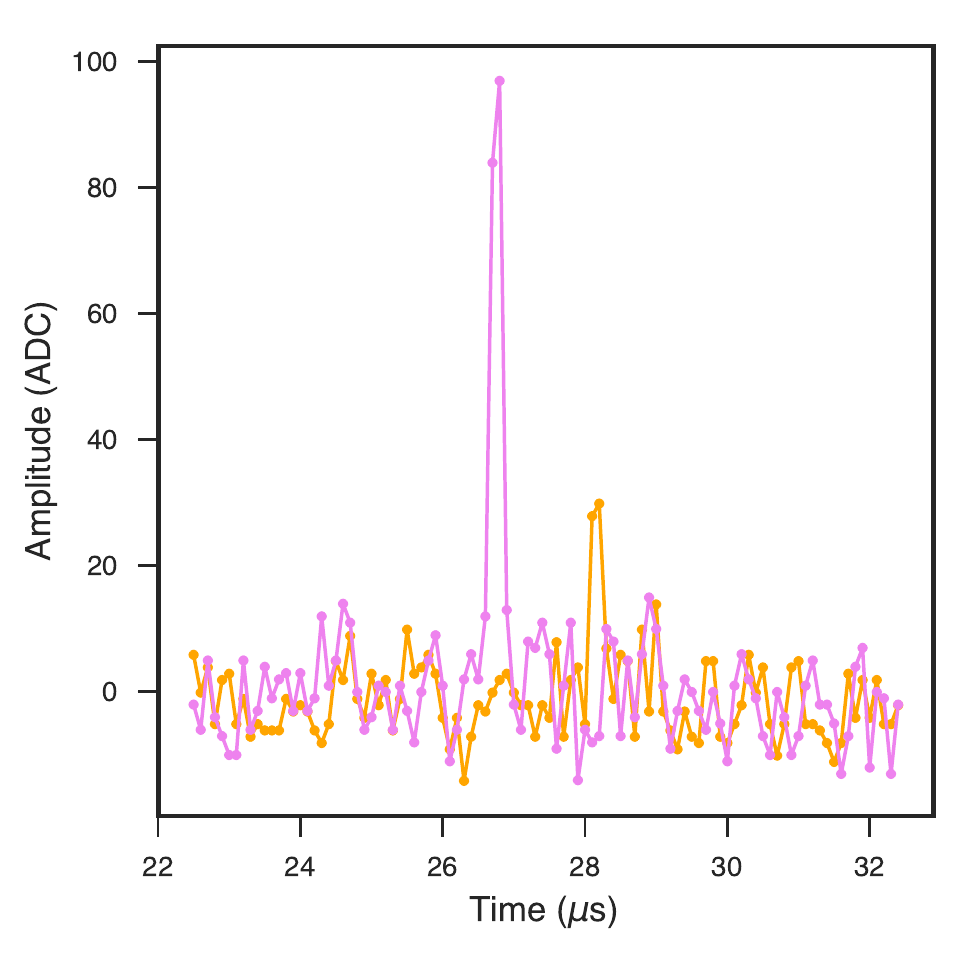}}
\caption[]{Upper panels (a and b) show the camera image of a triggered EAS-like event, (a) at $G{=}5000$ and (b) at nominal gain. The colour scale indicates the peak time in the individual pixels to illustrate the time gradient within the camera. Lower panels (c and d) show the traces of the upper events in two selected pixels (similar colour as with black dots marked pixels in upper panels).}
\label{fig:EAS_FDEyeDisplay}
\end{figure*}

In addition to the performance tests, 40 T3 triggers\footnote{The FD camera trigger is a three--level--trigger: A T1 is generated when the signal of a single pixel is above a given threshold during a certain time period. The T2 looks for valid patterns in a given, wider time window, and the T3 applies further timing constraints to find patterns. Detailed explanations are given in \cite{Abraham:2009pm}.} were collected during $\sim$28 hours of data taking with Los Leones 1 operated at three different gain settings depending on the NSB conditions. Eight of the T3 trigger events could be classified as shower candidates. Two of those were detected at a gain of $G=3000$ (factor 16.6 lower than nominal), five at a gain of $G=5000$ (factor 10 lower), and one at nominal gain. It should be emphasised that the measurement campaign with Los Leones 1 presented in this work was undertaken first of all to prove the principle of concept and to verify lab results on site, and not (yet) to generate physical data. This is why neither a detailed energy reconstruction nor a trigger optimisation for fluorescence measurements at reduced gains were performed beforehand or at the time of the analysis. This is planed for the future within the scope of the Auger upgrade \cite{Engel:2015ibd}. Fig.~\ref{fig:EAS_FDEyeDisplay} shows camera images and traces for two events, one detected at $G=5000$ and the other at nominal gain. In both cases, the signal peak is high enough to be identified and separated from baseline fluctuations.

\section{Conclusion and outlook}
\label{conclusion}
%% conclusion.tex %%
%
%----------------------------------------
%
The main limitation of the duty cycle of existing PMT based cameras results from fast PMT ageing when exposed to a high background light, e.g.~high NSB level. To counteract this, the PMT gain can be reduced for periods of increased background light conditions. Therefore, a PMT test stand was designed and built to investigate the PMT performance and ageing under different operation conditions and for different gain settings. It is also usable for many other measurements of this kind in future.

Nine spare PMTs of type Photonis XP3062 as used in the Auger FD cameras were characterised and tested. Full measurement cycles at increased continuous background light conditions and reduced PMT gains were emulated in lab measurements showing that lowering the gain by a factor of 10 is reasonable for the operation at increased NSB. The ageing was experimentally verified: The overall loss of sensitivity shows no dependence on the initial gain even though differences in the recovery from breaks are observed for a PMT with an initial gain of $G=500$, an order of magnitude lower than the gain proposed to be used for increased-NSB measurements.

Furthermore, one fluorescence telescope of the Pierre Auger Observatory was operated -- for the first time -- at reduced gain proving that a measurement mode with switching gain settings throughout a night according to changing NSB conditions can be implemented and still deliver scientifically valuable data. It can be anticipated that in case of the Auger FD, the duty cycle could be increased from 15\% to a maximum value of 21\%, or even from 19\% to 29\% if not including the expected reduction due to bad weather, power cuts, and malfunctions.

Although this study was done using PMTs of a specific type (Photonis XP3062) for the use in cameras detecting atmospheric fluorescence light from EASs, the results of this study are transferable to other experiments with cameras exposed to continuous background light being equipped with PMTs of similar types such as used in imaging atmospheric Cherenkov telescope cameras.

\section*{Acknowledgements}
\label{ackn}
%% ackn.tex %%
%
%----------------------------------------
We gratefully acknowledge the support by the Pierre Auger Observatory providing us with spare PMTs for testing and giving us the opportunity to perform on-site measurements.

Furthermore, we thank Michael Unger for providing us with results from Monte-Carlo simulation.

\section*{References}
\bibliographystyle{elsarticle-num-names}
\bibliography{mybibfile}

%\end{linenumbers}
\end{document}